# A Practical Grid-Partitioning Method Considering the Dynamic VAR Response of Power Grid under Contingency Set

*Wenlu Zhao\**

*\*Tsinghua University, Beijing, China*

## Abstract

Due to the increasing load demand, the operating state of modern power systems are become closing to their critical levels. Moreover, the modern power systems are integrated with a large amount of fast-response dynamic elements, so the short-term voltage stability (STVS) issues should be taken seriously. In order to improve the STVS of power systems, we can optimize the dynamic VAR reserve under contingency set. However, the scale of actual power grids is usually large, so it is hard to solve the overall optimization problem directly. To reduce the difficulty of solving the problem, a practical grid-partitioning method considering the dynamic VAR response of power grid under contingency set is proposed in this paper, to decompose the overall problem into several sub-problems. The buses in each region are similar in terms of voltage response to dynamic VARs and contingencies. The effectiveness of the proposed method is verified based on a region power grid model of China.

## Nomenclature

**A. Notation**

VR    Voltage response
SC    Severe contingency
RSC   Representative severe contingency

## 1  Introduction

Voltage stability is a key factor to ensure the steady and safe operation of power systems, so the voltage stability issues should be taken seriously. The IEEE joint task force on stability terms and definitions have published a report on the definition and classification of power system stability [1]. In the report, the voltage stability was divided into large-disturbance voltage stability and small-disturbance voltage stability in terms of the amplitude of disturbance, and was divided into long-term voltage stability and short-term voltage stability in terms of the time scale of disturbance.

Due to the increasing load demand, the operating state of modern power systems are become closing to their critical levels. If a contingency occurs in the power system, the voltage stability of the power system will be threatened. Moreover, the modern power systems are integrated with a large amount of fast-response dynamic elements, such as induction motors, renewable power generation equipment and HVDC converter stations. Therefore, the response of the power systems after contingencies will be complicated and will be rapidly changing, thus the short-term voltage stability (STVS) issues should be taken seriously.

In order to improve the STVS of power systems, we can optimize the dynamic VAR reserve under a set of anticipated contingencies. However, the scale of actual power grids is usually large, with many anticipated contingencies and dynamic VAR resources. So the overall optimization problem of the whole power grid is very complex, and is hard to solve directly. Fortunately, the voltage issues are usually regional. Therefore, we can try to decompose the overall problem into several sub-problems, thus decreasing the difficulty of solving the problem of optimizing dynamic VAR reserve.

Currently, most of the grid-partitioning researches focus on the static voltage stability issue [2]. In general, these methods are all aimed at steady-state voltage control. The sensitivity between bus voltage and reactive power are used to evaluate the similarity of buses in terms of voltage control, and further realizes the decoupling control of different regions in the power grid. However, the dynamic VAR response of power grids under contingency set are not considered in these methods, so there will be challenges while applying to the STVS issue.

In order to meet the requirements of the STVS issue, a practical grid-partitioning method considering the dynamic VAR response of power grid under contingency set is proposed in this paper. The content of this method is as follows. Firstly, a short-term voltage stability index ( *STVSI* ) is proposed to describe the STVS of a voltage trajectory. Then the representative severe contingencies (RSCs) in the contingency set can be filtered out. Secondly, the roughly grid-partition based on the similarity under contingencies can be carried out. Thirdly, an index for measuring the trajectory sensitivity of voltage trajectories to dynamic VARs ( *VQTSI* ) is proposed. Then the further grid-partition can be carried out based on the dynamic VAR response of power grid.

The coupling relationship of the buses with contingencies and dynamic VARs can be further obtained based on the proposed grid-partitioning method. When we study a certain region of the power system, we can only focus on the contingencies and dynamic VARs which are closely coupled with it. Therefore, we can decompose the overall problem into several sub-problems, thus simplifying the originally very complex issue.

The contents of this paper is as follows. In section 2, the grid-partitioning method considering dynamic VAR response of power grid under contingency set is proposed. In section 3, the effectiveness of the proposed method is verified based on a practical regional power grid of China. In section 4, some conclusions of this research is presented.

## 2 Dynamic VAR Response Grid-Partitioning Method

In this section, the method of grid-partitioning considering dynamic VAR response of power grid will be stated in detail. The method can be divided into three steps.

Firstly, filter out representative severe contingencies (RSCs). In practical operation, only a minority of the anticipated contingencies will threaten the STVS of the power system. Moreover, some of these minority contingencies have similar impact on the power system. Therefore, we can only focus on the representative ones in these minority contingencies in terms of STVS issue, thus can simplifying the issue while not significantly affecting the analysis result. In this paper, these minority contingencies are called as SCs. An index ($STVSI$) for evaluating the STVS of voltage trajectories is proposed, which will be used in filtering RSCs.

Secondly, roughly grid-partition considering voltage response (VR) under the RSCs. Since the voltage stability issues are regional, the contingencies usually only have a significant impact on part of the power system. It can be concluded that the SCs which should be noticed by the buses that close to each other are similar. Therefore, we can select the RSCs from the SCs, and then make a roughly grid-partition of the power system according to the similarity of VR to the RSCs.

Thirdly, further grid-partition according to the VR to dynamic reactive power resources. Since the proposed grid-partitioning method is trying to decouple the overall optimization problem to several sub-problems so as to decrease the difficulty to solve the problem, we need to guarantee that the buses in each partition have similar VR to dynamic VARs. However, the distribution of dynamic VARs differs from the distribution of RSCs, so the similarity of the VR of the buses in each rough partition cannot be guaranteed. Therefore, we need further grid-partition according to VR of the buses to dynamic VARs, based on the former rough grid-partitioning result. In this paper, an index for measuring the trajectory sensitivity of voltage trajectories to dynamic VARs ($VQTSI$) is proposed, which will be used for evaluating the similarity of the buses according to the VR to dynamic VARs.

### 2.1 Filter out representative severe contingencies (RSCs)

This section contains five parts, as shown in Table 1.

| Step | Contents |
| --- | --- |
| 1 | Define an index ($STVSI$) to evaluate the STVS of voltage trajectories |
| 2 | Scan the contingency set using time domain simulation |
| 3 | Filter out SCs utilizing $STVSI$ |
| 4 | Calculate the scope of each SCs utilizing $STVSI$ |
| 5 | Select RSCs from SCs utilizing $STVSI$ |

Table 1 Steps to filter out RSCs

The first step is to define an index ($STVSI$) to evaluate the STVS of voltage trajectories. In order to filter out SCs and further select RSCs, an index which can quantitatively describe the impact of contingencies to the bus voltage is required. Currently, the practical STVS criteria/index can only determine whether the bus voltage is stable, but cannot describe the relative degree of stability of the bus voltage [3-8]. So these criteria are not feasible to select the representative ones in the SCs. Therefore, the current practical criteria cannot fully show the similarity of the contingencies in terms of the response of bus voltage, and thus cannot fully simplify the overall STVS optimization problem.

The index proposed in this section is based on voltage trajectories. Before the definition of the $STVSI$, the definition of some parameters is shown in Table 2.

| Parameter | Definition |
| --- | --- |
| $v(t)$ | Voltage trajectory of a bus |
| $T_{end}$ | Time instant at the end of $v(t)$ |
| $T_{th}$ | Time duration threshold for continuous low voltage in $v(t)$ |
| $V_{th}$ | Low voltage threshold for $v(t)$ |

Table 2 Definition of some parameters for $STVSI$

In practice, if the time duration that $v(t)$ persists below $V_{th}$ after contingency exceed $T_{th}$, then the STVS of the bus is considered to be weak, and the corresponding contingency can be considered as a SC. For instance, according to the WECC Planning Standards [3-4], the voltage dip should not exceed 20% for more than 20 cycles at load buses under an event resulting in the loss of a single element. The definition of $STVSI$ is as follows:

In the definition, $T_{span,\max}$ is the length of the longest time duration that $v(t)$ persists below $V_{th}$, and $STVSI$ is the value of $T_{span,\max}$ divided by $T_{th}$. The range of $STVSI$ is $[0,+\infty)$. The contingency becomes more severe as $STVSI$ increases. When $STVSI$ equals to 1, the severity of the contingency is critical. The schematic of $STVSI$ is shown in Fig 1.

$$S_{Tspan} = \left\{ \{t_1, t_2\} \middle| \begin{array}{l} v(t_1) = v(t_2) = V_{th}, 0 \le t_1 < t_2 \le T_{end}; \\ \forall t \in [t_1, t_2], v(t) < V_{th} \end{array} \right\} \quad (1)$$

$$T_{span,\max} = \begin{cases} \max(t_2 - t_1), \{t_1, t_2\} \in S_{Tspan} & \text{if } S_{Tspan} \ne \emptyset \\ 0 & \text{if } S_{Tspan} = \emptyset \end{cases} \quad (2)$$

$$STVSI = T_{span,\max} / T_{th} \quad (3)$$

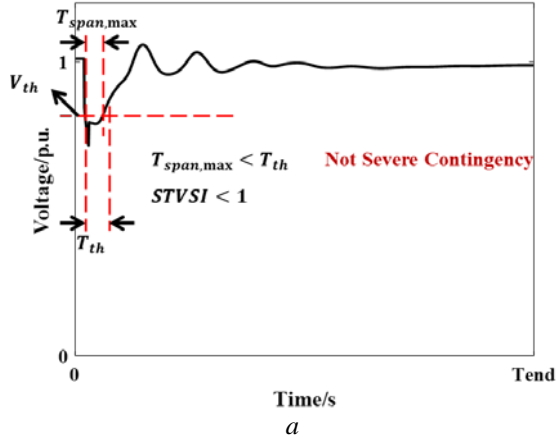

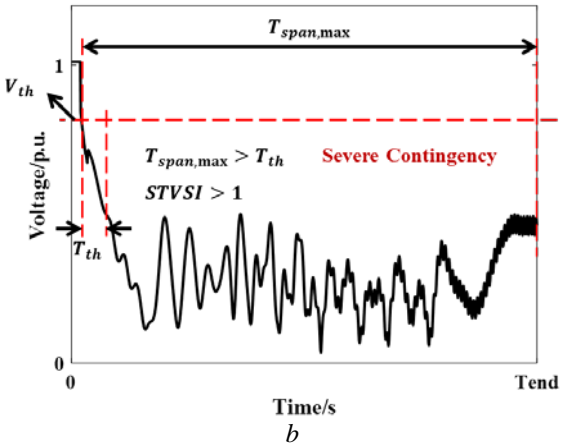

Fig 1 Schematic of $STVSI$
(*a*) $STVSI$ under a voltage stable case, (*b*) $STVSI$ under a voltage unstable case

It should be noted that $STVSI$ describes the VR of a bus to a contingency. If more than one bus or more than one contingency is considered, the $STVSI$ should be weighted or integrated into a matrix for analysis.

The second step is to scan the contingency set using time domain simulation. For the STVS issue of practical large scale power grids, the time domain simulation method is almost the only feasible method.

The numerical simulation software is PSASP (Power System Analysis Software Package), which is developed by CEPRI (China Electric Power Research Institute). This software is widely used in regional power grids of China. The $STVSI$ of voltage trajectories is required to be calculated after simulation, which will be used in following steps.

The third step is to filter out SCs utilizing $STVSI$. The SCs have a more severe impact on the power grid in terms of STVS compared to the not severe ones, so only the SCs will be considered in the following steps.

According to the definition of $STVSI$, the STVS of a bus is weak and the contingency is severe if the value of the index is greater than 1. In this paper, if the $STVSI$ of any bus in the power system is greater 1, then the corresponding contingency is considered as a SC. The mathematical expression is as follows:

$$S_{sFlt} = \left\{ Flt_i \middle| \max_{Bus_j \in S_{Bus}} \left( STVSI(Flt_i, Bus_j) \right) > 1, Flt_i \in S_{Flt} \right\} \quad (4)$$

$$S_{Flt} = \{Flt_1, ..., Flt_k\} \quad (5)$$

$$S_{Bus} = \{Bus_1, ..., Bus_n\} \quad (6)$$

In the expressions, $S_{sFlt}$ is the set of SCs, $S_{Flt}$ is the set of contingencies, $S_{Bus}$ is the set of Buses. The $STVSI$ is considered as a map from a contingency and a bus. The amount of contingencies is assumed to be $k$, and the amount of buses is assumed to be $n$.

The fourth step is to calculate the scope of each SCs utilizing $STVSI$. For a SC, the impact inside the scope is more severe in terms of STVS compared to the impact outside the scope, so only the buses in the scope of the SC will be considered in the following steps.

In this paper, the scope of a SC is defined as the set of buses whose $STVSI$ is greater than 1 under this SC. The mathematical expression is as follows:

$$S_{sFlt_i, Bus} = \left\{ Bus_j \middle| STVSI(sFlt_i, Bus_j) > 1, Bus_j \in S_{Bus} \right\} \quad (7)$$

$$sFlt_i \in S_{sFlt} \quad (8)$$

$$S_{Bus} = \{Bus_1, ..., Bus_n\} \quad (9)$$

In the expressions, $S_{sFlt_i, Bus}$ is the scope of SC $sFlt_i$.

The last step is to select RSCs utilizing $STVSI$. The RSCs can represent all SCs in terms of the impact on the STVS of the power grid, so only the RSCs will be considered in the following steps. In this paper, if the $STVSI$ of the buses under contingency $i$ in the scope of contingency $i$ are all smaller than those under contingency $j$, then the effect of contingency $i$ is considered to be totally represented by contingency $j$. If the effect of a contingency cannot be represented by any other contingency, then this contingency is defined as a representative contingency. The mathematical expression is as follows:

$$\text{if } \forall Bus_l \in S_{sFlt_i,Bus} \Rightarrow$$
$$STVSI(sFlt_i, Bus_l) < STVSI(sFlt_j, Bus_l) \quad (10)$$
$$\text{then } sFlt_j \in S_{rsFlt_i}$$

$$\text{if } S_{rsFlt_i} = \varnothing$$
$$\text{then } sFlt_i \in S_{rsFlt} \quad (11)$$

$$sFlt_i, sFlt_j \in S_{sFlt} \quad (12)$$

In the expressions, $S_{rsFlt_i}$ is the set of contingencies which can represent the effect of $sFlt_i$, $S_{rsFlt}$ is the set of RSCs.

## 2.2 Roughly grid-partition considering representative severe contingencies (RSCs)

This section contains two parts, as shown in Table 3.

| Step | Contents |
|---|---|
| 1 | For each bus, mark out the RSCs whose scope include the bus |
| 2 | Classify the buses according to the mark of RSCs |

Table 3 Steps to roughly grid-partition considering RSCs

The first step is marking out the RSCs corresponding to each bus. For each RSC, there is a corresponding scope in terms of the severity of STVS. Similarly, for each bus, there is a corresponding set of RSCs in terms of the severity of STVS. After marking, the buses can be roughly classified. Base on the RSCs and the scope of SCs, the task of marking is easy to accomplish. The mathematical expression is as follows:

$$S_{Bus_i,rsFlt} = \left\{ rsFlt_j \middle| Bus_i \in S_{rsFlt_j,Bus}, rsFlt_j \in S_{rsFlt} \in S_{sFlt} \right\} \quad (13)$$

In the expression, $S_{rsFlt_j,Bus}$ is the scope of RSC $rsFlt_j$.

The second step is grid-partitioning according to the mark of RSCs. The buses with the identical mark of RSCs are considered to be more similar compared to the buses with different mark of RSCs. Actually, the task of this step is classifying the buses according to the marked RSCs. After partitioning, the buses in each region are considered to be highly similar in terms of VR to contingency set. The mathematical expression is as follows:

$$S_{cBus} = \{C_1,...,C_m\}$$
$$s.t. \ \forall C_k, C_l \in S_{cBus} \Rightarrow C_k \cap C_l = \varnothing$$
$$C_1 \cup \cdots \cup C_m = S_{Bus} \quad (14)$$
$$\forall Bus_i, Bus_j \in C_k \Rightarrow S_{Bus_i,rsFlt} = S_{Bus_j,rsFlt}$$
$$\forall Bus_i \in C_k, Bus_j \in C_l, k \neq l \Rightarrow S_{Bus_i,rsFlt} \neq S_{Bus_j,rsFlt}$$

## 2.3 Further grid-partition considering dynamic VARs

This section consists four parts, as shown in Table 4.

| Step | Contents |
|---|---|
| 1 | Define an index for measuring the trajectory sensitivity of voltage trajectories to dynamic VARs ( *VQTSI* ) |
| 2 | Construct the distance matrix between the buses in terms of the VR to a certain dynamic VAR under a certain RSC |
| 3 | Construct the distance matrix between the buses in terms of the VR to all corresponding dynamic VARs under all corresponding RSCs |
| 4 | Visualize the distance matrix of the buses into two-dimensional data points, then cluster the data points |

Table 4 Steps to further grid-partition considering dynamic VARs

The first step is defining an index to measure the trajectory sensitivity of voltage trajectories to dynamic VARs, which will be called as *VQTSI* in this paper. The VR of bus will change if the dynamic VAR reserve changes. However, the change of VR is a trajectory, while the change of dynamic VAR reserve is a scalar. To quantify the impact of dynamic VARs on the voltage trajectories, an expression to quantify the change of voltage trajectories is firstly defined, then the expression can be used to quantify the sensitivity of dynamic VAR reserve to voltage trajectories. The definition of *VQTSI* is as follows:

$$\Delta Vt = S_2 - S_1 \quad (15)$$
$$VQTSI = \Delta Vt / \Delta Q \quad (16)$$
$$S_1 = \int_0^{T_{end}} \max(0, V_{10} - v_1(t)) dt \quad (17)$$
$$S_2 = \int_0^{T_{end}} \max(0, V_{20} - v_2(t)) dt \quad (18)$$
$$V_{10} = v_1(t)|_{t=0}, V_{20} = v_2(t)|_{t=0} \quad (19)$$
$$\Delta Q = Q_2 - Q_1 \quad (20)$$

In the expression, $S_{cBus}$ is the set of grid-partitioning results, each element of $S_{cBus}$ is a set of buses.

In the definition, $\Delta Vt$ quantifies the change of voltage trajectories, *VQTSI* is the defined trajectory sensitivity index. The variables whose subscript contains 1 are the original ones, while the variables whose subscript contains 2 are modified ones. $v_1(t)$ and $v_2(t)$ are voltage trajectories of buses, $V_{10}$ and $V_{20}$ are the initial bus voltage, $Q_1$ and $Q_2$ are the capacity of dynamic VAR reserve. The range of *VQTSI* is $(-\infty, +\infty)$. The greater is the absolute value of $VQTSI$, the greater is the impact of dynamic VARs on voltage trajectories. If $VQTSI$ equals to 0, the voltage trajectory will not be affected by the dynamic VAR. The schematic of $\Delta Vt$ is shown in Fig 2.

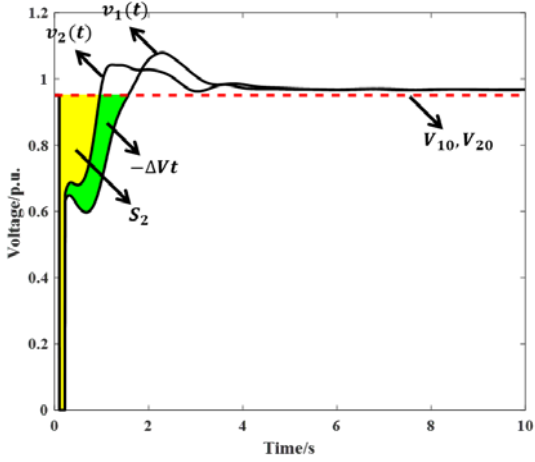

Fig 2 Schematic of $\Delta Vt$

In Fig 2, the area of yellow region is $S_2$, the area of yellow region plus green region is $S_1$. So the area of green region is the opposite value of $\Delta Vt$.

It should be noticed that this sensitivity is different from the commonly used trajectory sensitivity. In the commonly used trajectory sensitivity, the sensitivity trajectories are firstly calculated, then an index to describe the sensitivity is defined. However, in this sensitivity, an index to describe the change trajectories is firstly defined, then the sensitivity of the index is calculated. Additionally, the $VQTSI$ describes the VR of a bus to a contingency and a dynamic VAR. If more than one bus, or more than one contingency, or more than one dynamic VAR is considered, the $VQTSI$ should be weighted or integrated into a matrix for analysis.

The second step is constructing the distance matrix between the buses in terms of the VR to a certain dynamic VAR under a certain RSC. In order to comprehensively consider the VR of all buses to a certain dynamic VAR under a certain RSC, the $VQTSI$ corresponding to each bus are integrated into a matrix. The constructed distance matrix can obviously describe the similarity of buses corresponding to a dynamic VAR under a certain RSC in terms of VR. The mathematical expression is as follows:

In the expressions, $d^{Flt_p, VAR_q}$ is the distance matrix corresponding to RSC $Flt_p$ and dynamic VAR $VAR_q$, $d_{i,j}^{Flt_p, VAR_q}$ is the distance between the $VQTSI$ of bus $i$ and bus $j$ corresponding to RSC $Flt_p$ and dynamic VAR $VAR_q$, $VQTSI_i^{Flt_p, VAR_q}$ is the $VQTSI$ of bus $i$ corresponding to RSC $Flt_p$ and dynamic VAR $VAR_q$.

$$d^{Flt_p, VAR_q} = \begin{bmatrix} 0 & d_{1,2}^{Flt_p, VAR_q} & \cdots & d_{1,n-1}^{Flt_p, VAR_q} & d_{1,n}^{Flt_p, VAR_q} \\ d_{2,1}^{Flt_p, VAR_q} & 0 & \cdots & d_{2,n-1}^{Flt_p, VAR_q} & d_{2,n}^{Flt_p, VAR_q} \\ \vdots & \vdots & \ddots & \vdots & \vdots \\ d_{n-1,1}^{Flt_p, VAR_q} & d_{n-1,2}^{Flt_p, VAR_q} & \cdots & 0 & d_{n-1,n}^{Flt_p, VAR_q} \\ d_{n,1}^{Flt_p, VAR_q} & d_{n,2}^{Flt_p, VAR_q} & \cdots & d_{n,n-1}^{Flt_p, VAR_q} & 0 \end{bmatrix} \quad (21)$$

$$d_{i,j}^{Flt_p, VAR_q} = \left| VQTSI_i^{Flt_p, VAR_q} - VQTSI_j^{Flt_p, VAR_q} \right| \quad (22)$$

The third step is constructing the overall distance matrix between the buses in terms of the VR to all corresponding dynamic VARs under all corresponding RSCs. In order to comprehensively consider the VR of the buses to all corresponding dynamic VARs and RSCs, the distance matrices obtained from the previous step will be weighted average. The weighted distance matrix can comprehensively describe the similarity of buses corresponding to all relevant dynamic VARs and RSCs in terms of VR. The mathematical expression is as follows:

$$d = \sum_{Flt_p=1}^{m} \sum_{VAR_q=1}^{l} \omega_{Flt_p} \omega_{VAR_q} d^{Flt_p, VAR_q} \quad (23)$$

$$\sum_{Flt_p=1}^{m} \omega_{Flt_p} = 1, \omega_{Flt_p} \geq 0 \quad (24)$$

$$\sum_{VAR_q=1}^{l} \omega_{VAR_q} = 1, \omega_{VAR_q} \geq 0 \quad (25)$$

In the expressions, $d$ is the weighted distance matrix, $\omega_{Flt_p}$ and $\omega_{VAR_q}$ are the weight of RSC $Flt_p$ and dynamic VAR $VAR_q$ respectively. The number of RSC and dynamic VAR is assumed to be m and l respectively.

It should be noticed that for each dynamic VAR, we assume that it is related to all buses. For each RSC, the weight is set to be 0 if a bus is outside its scope.

The fourth step is visualizing the distance matrix of the buses into two-dimensional data points, then cluster the data points using hierarchical clustering methods. The VR of the buses to dynamic VARs under RSCs are required to be similar, so the radius of clusters must be small enough. The bottom-up hierarchical clustering method is used in this paper, because the radius of clusters can be monitored during clustering process. However, the distance matrix cannot be directly used in the hierarchical clustering, so the distance matrix is converted into data points. In this paper, the Multi-Dimensional Scaling (MDS) method is utilized to map the distance matrix into two-dimensional data points. A brief introduction to the hierarchical clustering method and the MDS method is as follows.

The hierarchical clustering is a method to build a hierarchy of clusters. Methods of hierarchical clustering can generally be divided into two types [9]: bottom-up approach and top-down approach.

In the bottom-up method, each data point starts as a cluster, then pairs of clusters merged to one cluster while moves up the hierarchy. In the top-down method, all data points start as a cluster, then split it into pairs of clusters while moves down the hierarchy. The results of hierarchical clustering are usually presented in dendrogram. An example of dendrogram is shown as Fig 3.

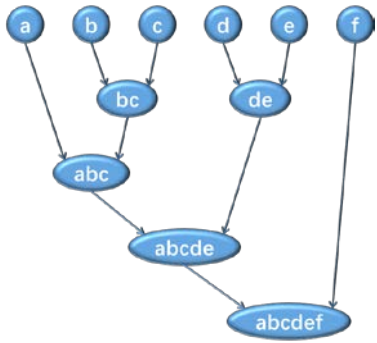

Fig 3 Schematic of a dendrogram

The Multidimensional scaling (MDS) is a method to visualize the similarity of samples in a dataset [10]. Particularly, the information stored in a distance matrix can be displayed through this method. An MDS algorithm aims to place each sample in N-dimensional space while preserving the distances between samples as well as possible. If N=2, a scatterplot can be made after the MDS process. An example of the MDS result is shown as Fig 4.

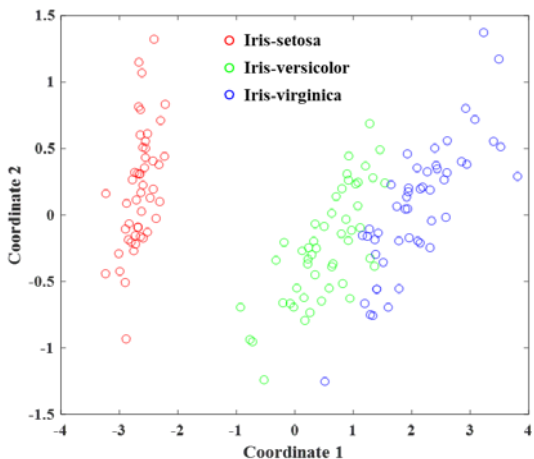

Fig 4 Example of a 2-dimensional MDS result based on the Iris flower data set

In Fig 4, the result of a 2-dimensional MDS process based on the Iris flower data set is presented. The Iris flower data set is a multivariate data set introduce by Ronald Fisher as an example of linear discriminant analysis [11]. This data set is a typical test case for many classification approaches [12]. The three types of flowers in the Iris data set are marked in red, green and blue respectively in Fig 4. The three colour samples are obviously clustered, indicates that the MDS method can preserve the similarity between samples while placing the samples in a 2-dimensional space.

**2.4 Summary**

In this section, the main content of the proposed dynamic VAR response grid-partitioning method is summarized, and a diagram to show the process of grid-partition is shown as Fig 5.

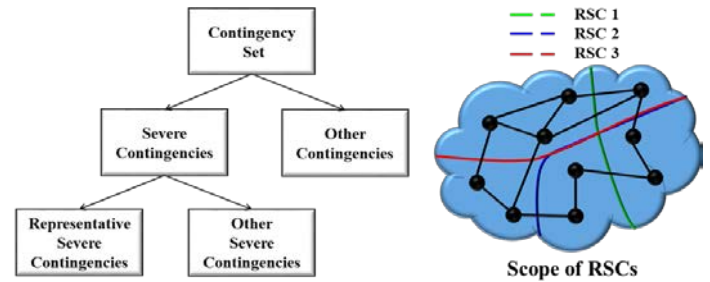

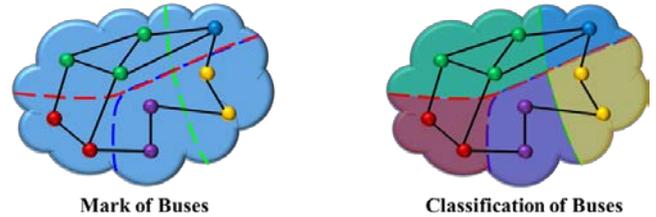

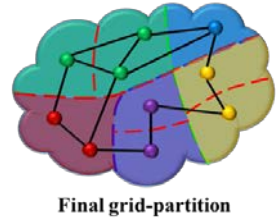

Fig 5 Process of the proposed grid-partitioning method

## 3 Case Studies

The case studies are based on a regional power grid model of China, which contains 3,000+ buses, 4,000+ AC lines, 400+ generators and 7 HVDC lines. This power grid receives large amount of electrical power from external, so it is lack of internal power generation. If the HVDC line is outage, the grid will be risky to voltage collapse. Therefore, this power grid is weak in terms of STVS. Additionally, the proportion of induction motors is high in load. In typical operating modes in summer, the proportion of induction motors exceeds 50%. Therefore, the response of the grid after contingencies will be complicated and will be rapidly changing.

To improve the STVS of this grid, optimizing dynamic VAR reserve under a set of anticipated contingencies is a feasible way. However, the scale of this grid is too large, so it is hard to solve the overall optimization problem directly. In order to decrease the difficulty of solving the optimization problem, we can decouple the grid into several parts, thus decomposing the overall problem into several sub-problems. Considering that the main concern of the grid is STVS issue, and the response of the grid after contingencies is complicated and rapidly changing, thus it is reasonable to use the proposed dynamic VAR response grid-partitioning method on this grid.

This section contains two parts. In the first part, the detailed grid-partitioning process will be presented. In the second part, the comparison of VR of buses in a same region and buses in different regions to contingencies and dynamic VARs will be presented to verify the effectiveness of the proposed method

### 3.1 An example of the grid-partitioning process

This case is based on a typical operating mode in summer. The contingency set contains 682 anticipated contingencies. All of the generators, Static VAR Compensators (SVCs) and Static Synchronous Compensators (STATCOMs) are considered as dynamic VAR resources.

After the definition of $STVSI$, the next step is to filter out SCs. In this case, the severity of the contingencies is determined by the $STVSI$ of the 500kV buses. The result of filtering out SCs is shown as Fig 6.

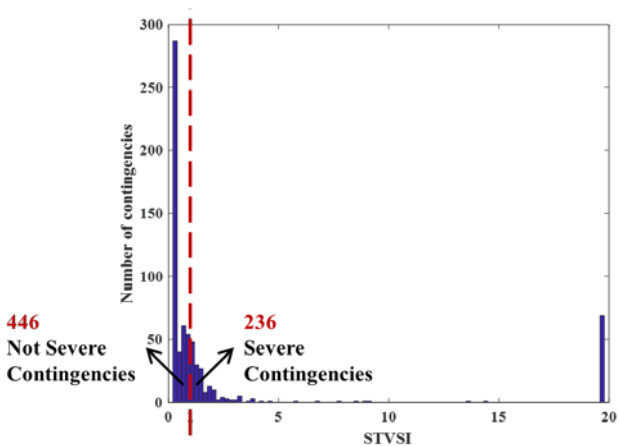

Fig 6 Result of filtering out SCs

From Fig 6, only 236 of the 682 anticipated contingencies are severe. The proportion of SCs in anticipated contingencies is less than 35%, so it is significant to filter out SCs.

After filtering out SCs, the next step is to select the RSCs from SCs. The result of RSCs is shown as Table 5.

| RSC ID | Number of SC represented | RSC ID | Number of SC represented | RSC ID | Number of SC represented |
|---|---|---|---|---|---|
| 1 | 30 | 14 | 22 | 27 | 2 |
| 2 | 35 | 15 | 24 | 28 | 2 |
| 3 | 37 | 16 | 24 | 29 | 2 |
| 4 | 39 | 17 | 24 | 30 | 2 |
| 5 | 19 | 18 | 24 | 31 | 2 |
| 6 | 32 | 19 | 152 | 32 | 2 |
| 7 | 32 | 20 | 24 | 33 | 2 |
| 8 | 34 | 21 | 24 | 34 | 5 |
| 9 | 32 | 22 | 7 | 35 | 2 |
| 10 | 4 | 23 | 134 | 36 | 2 |
| 11 | 9 | 24 | 16 | 37 | 2 |
| 12 | 8 | 25 | 16 | | |
| 13 | 7 | 26 | 6 | | |

Table 5 Result of RSCs

From Table 5, only 37 of the 236 SCs are representative. The proportion of RSCs in SCs is less than 16%, so it is significant to select RSCs.

After selecting RSCs, the next step is to mark the buses with RSCs and roughly grid-partition according to the marks. The result of roughly grid-partition is shown as Table 6

| Region ID | Number of 500kV buses | Region ID | Number of 500kV buses | Region ID | Number of 500kV buses |
|---|---|---|---|---|---|
| 1 | 48 | 16 | 3 | 31 | 1 |
| 2 | 1 | 17 | 2 | 32 | 1 |
| 3 | 21 | 18 | 1 | 33 | 1 |
| 4 | 13 | 19 | 1 | 34 | 1 |
| 5 | 1 | 20 | 1 | 35 | 1 |
| 6 | 2 | 21 | 2 | 36 | 1 |
| 7 | 1 | 22 | 3 | 37 | 1 |
| 8 | 6 | 23 | 7 | 38 | 1 |
| 9 | 1 | 24 | 2 | 39 | 1 |
| 10 | 97 | 25 | 7 | 40 | 1 |
| 11 | 16 | 26 | 12 | 41 | 2 |
| 12 | 5 | 27 | 7 | 42 | 1 |
| 13 | 1 | 28 | 1 | 43 | 1 |
| 14 | 9 | 29 | 1 | | |
| 15 | 2 | 30 | 3 | | |

Table 6 Result of roughly grid-partition

From Table 6, the whole power grid is divided into 43 regions. However, the size of some regions is too large. For instance, the region 10 contains 97 500kV buses. The next step is to further partition on these oversized regions considering the VR to dynamic VARs. The result of further grid-partition on region 10 is shown as Fig 7 and Table 7.

From Fig 7(b), there is a turning point when the number of clusters is 7, so region 10 of the previous step is further divided into 7 sub-regions. From Fig 7(c), the scales of the clusters are similar, and the data points in each cluster are gathered together, which partially verifies the effectiveness of the proposed method. From Table 7, the largest sub-region contains 46 500kV buses. The scales of the sub-regions are moderate, thus can guarantee the similarity of buses in each sub-region in terms of VR to dynamic VARs.

Fig 7 Result of further grid-partition on region 10
(*a*) Dendrogram of the clustering process, (*b*) Relative merging distance during clustering process, (*c*) Scatterplot of the 2-dimensional MDS result

| Sub-region ID | Number of 500kV buses | Sub-region ID | Number of 500kV buses | Sub-region ID | Number of 500kV buses |
|---|---|---|---|---|---|
| 1 | 4 | 4 | 15 | 7 | 46 |
| 2 | 14 | 5 | 5 | | |
| 3 | 8 | 6 | 5 | | |

Table 7 Result of further grid-partition on region 10

### 3.2 Effectiveness of the grid-partitioning method

This case is also based on the typical operating mode in summer. Additionally, the capacity of some dynamic VARs are modified, for observing the VR to dynamic VARs of the buses in each region.

The first part is to verify the similarity of VR under contingencies in each region. The result is shown as Fig 8.

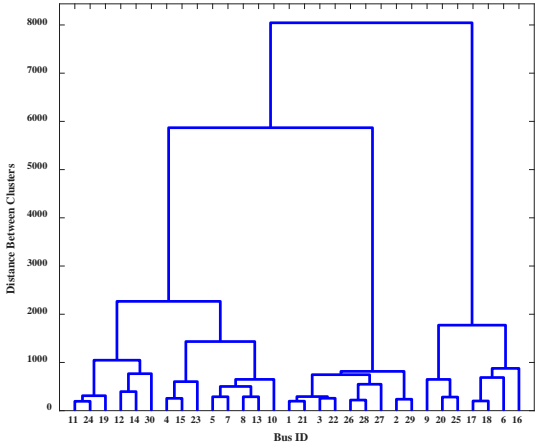

*a*

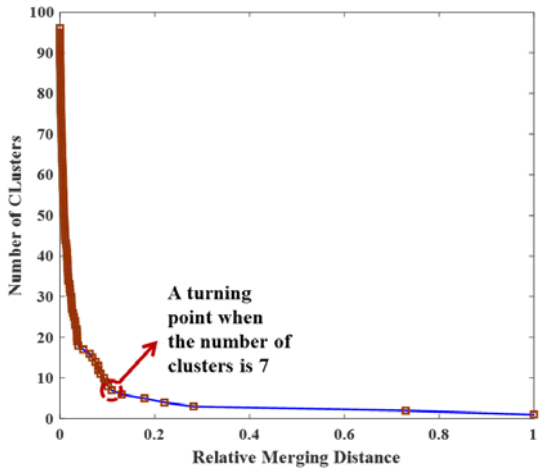

*b*

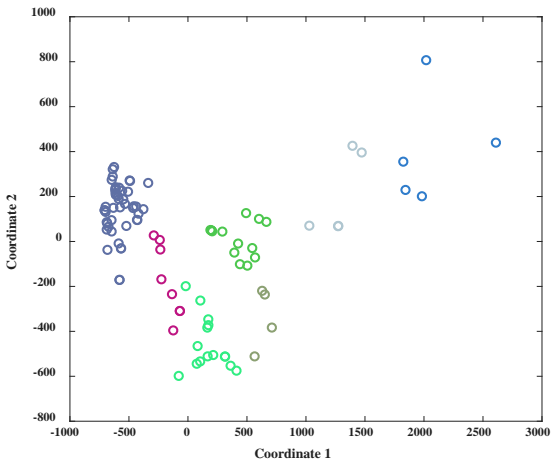

*c*

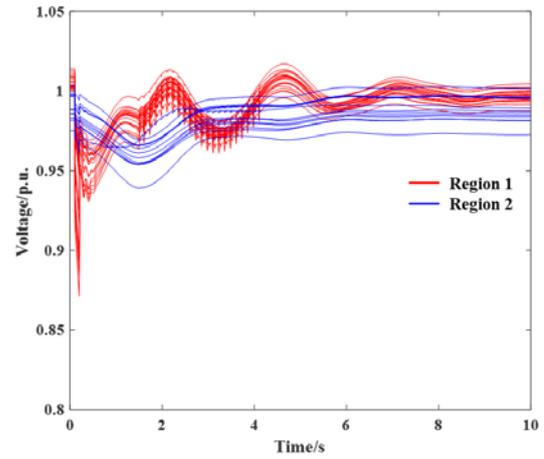

*a*

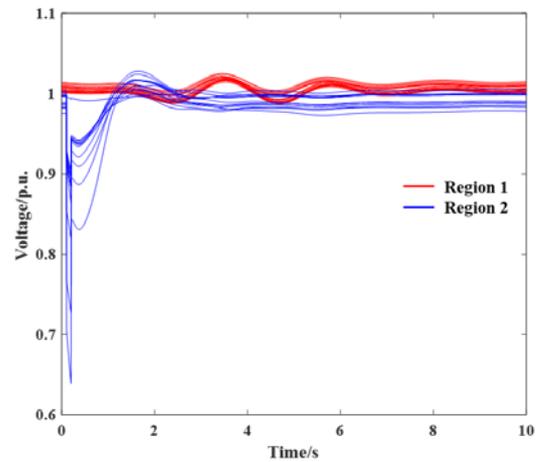

*b*

Fig 8 Voltage trajectories of the buses in two regions under two contingencies
(*a*) Voltage trajectories of the buses in two regions under one contingency, (*b*) Voltage trajectories of the buses in two regions under another contingency

From Fig 8, the voltage trajectories in each region are all similar under both of the contingencies, while the voltage trajectories of different regions are significantly different. Therefore, similarity of VR under contingencies in each region is verified.

The second part is to verify the similarity of VR to dynamic VARs in each region. The result is shown as Fig 9.

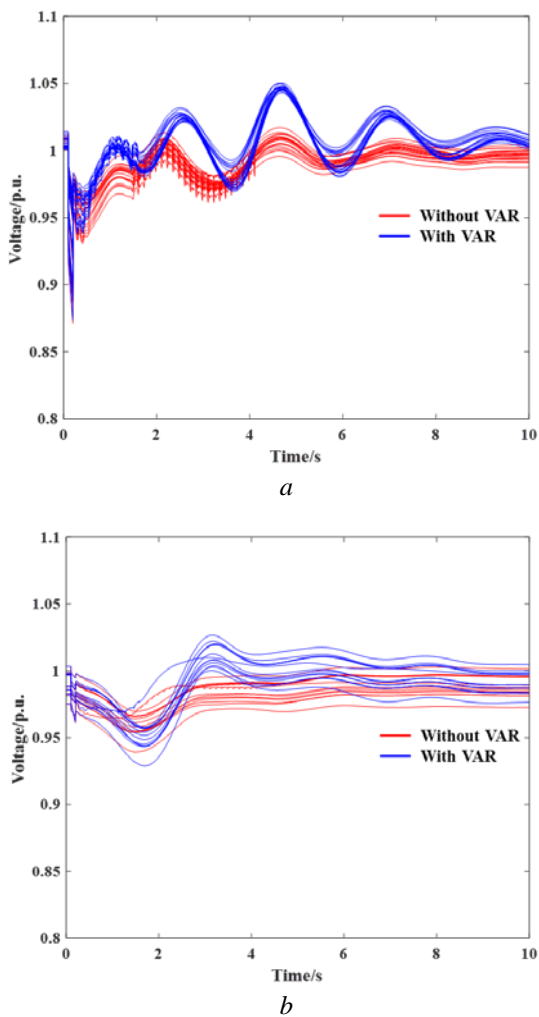

Fig 9 Voltage trajectories of the buses in a region under two configuration of dynamic VARs
(*a*) Voltage trajectories of the buses in one region under two configuration of dynamic VARs, (*b*) Voltage trajectories of the buses in another region under two configuration of dynamic VARs

From Fig 9, the voltage trajectories in a region are all similar under both configuration of dynamic VARs. Therefore, the similarity of VR to dynamic VARs in a region is verified.

## 4 Conclusion

A grid-partitioning method considering the dynamic VAR response of power grid is proposed in this paper, in order to decrease the complexity of the optimization problem of the whole power grid for improving STVS. Compared to most of the currently used grid-partitioning method, the proposed method considers the VR of power grid after contingencies, so it is more suitable in terms of STVS issue. After grid-partition, it can be ensured that the buses in each region are similar in terms of the VR to contingencies and dynamic VARs. Based on a regional power grid model of China, the effectiveness of the proposed method was verified.

In the future research, the dynamic VARs and contingencies which closely coupled to each region will be specified. Then the overall optimization problem will be simplified, and thus it will be easier to solve.